\documentclass[a4paper,UKenglish,cleveref, autoref, thm-restate]{lipics-v2021}

\title{Towards an Automated Reasoning Tool for Complexity Analysis of Automated Reasoners}

\author{Louis Rustenholz}{Universidad Politécnica de Madrid (UPM), Spain \and IMDEA Software Institute, Spain}{louis.rustenholz@imdea.org}{https://orcid.org/0000-0002-1599-2431}{}
\author{Manuel V. Hermenegildo}{Universidad Politécnica de Madrid (UPM), Spain \and IMDEA Software Institute, Spain}{}{https://orcid.org/0000-0002-7583-323X}{}
\author{Pedro López-García}{Spanish Council for Scientific Research (CSIC), Spain \and IMDEA Software Institute, Spain}{}{https://orcid.org/0000-0002-1092-2071}{}
\author{Alessio Mansutti}{IMDEA Software Institute, Spain}{}{https://orcid.org/0000-0002-1104-7299}{}
\author{Félix Ridoux\footnote{Work performed under previous IMDEA affiliation.}}{Computer Science Laboratory of Sorbonne University (LIP6), France \and STMicroelectronics, Crolles, France}{}{https://orcid.org/0009-0002-6312-0483}{}
\author{Niki Vazou}{IMDEA Software Institute, Spain}{}{https://orcid.org/0000-0003-0732-5476}{}
\authorrunning{L. Rustenholz et al}
\Copyright{Louis Rustenholz, Manuel V. Hermenegildo, Pedro López-García, Alessio Mansutti, Félix Ridoux, and Niki Vazou}



\category{Extended abstract}

\acknowledgements{
    Partially funded by MICIU projects CEX2024-001471-M \emph{María de
      Maeztu}, %
    by the European Union ERC GA 101039196 CRETE,
    and by the European Union MSCA GA 101154447 NEAT.
    We also thank the anonymous reviewers for their
    useful and constructive feedback.}

\EventEditors{Florian Frohn, and Étienne Payet}
\EventNoEds{21}
\EventLongTitle{21st International Workshop on Termination (WST 2026)}
\EventShortTitle{WST 2026}
\EventAcronym{WST}
\EventYear{WST 2026}
\EventDate{July 25, 2026}
\EventLocation{Lisbon, Portugal}
\EventLogo{}
\SeriesVolume{}
\ArticleNo{N}

\newtheorem*{example*}{Example}

\usepackage{xspace}

\usepackage{amsmath,amssymb,amsfonts}
\usepackage{mathtools}
\usepackage{stmaryrd}

\usepackage{galois}

\usepackage{tikz}
\usetikzlibrary{arrows}
\usetikzlibrary{arrows.meta}
\usetikzlibrary{calc}
\usetikzlibrary{intersections}
\usetikzlibrary{matrix}
\usetikzlibrary{patterns}
\usetikzlibrary{shapes}
\usetikzlibrary{shapes.geometric}
\usetikzlibrary{snakes}
\usetikzlibrary{positioning}
\usetikzlibrary{3d}
\usetikzlibrary{spy}
\usetikzlibrary{decorations.text}
\usetikzlibrary{decorations.pathmorphing}
\usetikzlibrary{decorations.markings}
\usetikzlibrary{babel}
\usetikzlibrary{plotmarks}
\usetikzlibrary{fit}
\usetikzlibrary{backgrounds}

\tikzset{>=Stealth}
\tikzset{every path/.style={line width=0.6pt}}

\tikzset{%
  toolstep/.style n args={2}{%
    draw,
    fill = white,
    minimum width  = #1,
    minimum height = #2,
    inner sep = 3.5pt,
    align = center,
    rounded corners = 2pt,
    line width = 0.6pt
  },
  toolstep/.default={1.0cm}{0.8cm}
}

\hideLIPIcs
\nolinenumbers

\usepackage{multido}
\usepackage{enumitem}
\usepackage{float}

\newcommand{\newextmathcommand}[2]{%
    \newcommand{#1}{\ensuremath{#2}\xspace}
}

\newcommand{\tool}{our tool}%
\newcommand{\IR}{IR\xspace}

\NewCommandCopy{\rawPhi}{\Phi}
\renewcommand{\Phi}{\mathrm{\rawPhi}}

\newextmathcommand{\fsol}{f_{sol}}
\newextmathcommand{\fcand}{\hat{f}}

\newextmathcommand{\nn}{\mathbb{N}}
\newextmathcommand{\nnp}{\Nat_+}
\newextmathcommand{\zz}{\mathbb{Z}}
\newextmathcommand{\qq}{\mathbb{Q}}
\newextmathcommand{\rr}{\mathbb{R}}
\newextmathcommand{\bb}{\mathbb{B}}

\newextmathcommand{\rri}{\overline{\rr}}

\newextmathcommand{\ListInt}{\mathtt{ListInt}}
\newextmathcommand{\Cons}{\mathtt{Cons}}
\newextmathcommand{\Nil}{\mathtt{Nil}}

\newextmathcommand{\sconcat}{\mathtt{f}}

\newextmathcommand{\cardm}{\mathtt{card}}
\newextmathcommand{\len}{\mathtt{len}}

\newcommand{\semop}[1]{\ensuremath{\left\llbracket#1\right\rrbracket_\Phi}\xspace}

\newcommand{\asemop}[1]{\ensuremath{\left\llbracket#1\right\rrbracket_{\Phi}^\sharp}\xspace}

\newextmathcommand{\Data}{\mathrm{Data}}
\newextmathcommand{\Metrics}{\mathcal{M}}

\newextmathcommand{\ii}{\mathcal{I}}
\newextmathcommand{\pp}{\mathcal{P}}

\newextmathcommand{\End}{\mathrm{End}}

\newextmathcommand{\Prog}{\mathtt{P}}
\newextmathcommand{\Funs}{\mathcal{F}}

\newextmathcommand{\pleq}{\mathbin{\dot\leq}}

\DeclareMathOperator{\lfp}{\mathrm{lfp}}
\DeclareMathOperator{\gfp}{\mathrm{gfp}}

\newextmathcommand{\threeexptime}{\textup{\textsc{3ExpTime}}}
\newextmathcommand{\tower}{\textup{\textsc{Tower}}}
\newextmathcommand{\poly}{\text{poly}}

\newextmathcommand{\card}{\#}

\newextmathcommand{\parhom}{\textit{hom}}
\newextmathcommand{\parmod}{\textit{mod}}
\newextmathcommand{\parconst}{\textit{const}}

\begin{document}

\maketitle

\begin{abstract}
We present the theory underpinning a complexity analysis tool (currently under development) that aims to automate tedious parts of the analysis of complex algorithms originating in the field of automated reasoning. Examples are given by super-exponential quantifier elimination procedures in real and integer arithmetic.

Our tool implements the following pipeline:
\vspace*{-4pt}
\begin{enumerate}
\item Together with the algorithm to be analysed, the user (an expert, e.g.\ the algorithm designer) can provide key metrics to track and lemmas to guide and improve the analysis. In pen-and-paper proofs, these correspond to the ``non-tedious'' and ``creative'' parts of the complexity analysis, that require human ingenuity.

\item The second step consists in the extraction of (generalised) recurrence equations. Here, we rely on a novel higher-order abstract interpretation technique, based on
operator semantics. It enables (optimal) abstract compilation of symbolic programs into different kinds of purely numerical recursive representations, such as recurrence equations on interval-valued functions or numerical logic programs.

\item Finally, our tool solves the recurrence equations. We propose going beyond the direct use of computer algebra systems (CAS) by employing pre/postfixpoint-based techniques to discover and verify candidate bounds on the solutions. This approach, in turn, leverages recent progress in SMT solvers, and could benefit from techniques originating in termination-analysis research.
\end{enumerate}
\vspace{-4pt}
\end{abstract}
\section{Introduction}
\label{sec:intro}

In current practice, the complexity analysis of algorithms in computational
logic is carried out almost exclusively by hand.
Such analyses typically require pen-and-paper proofs spanning tens of pages,
involving the derivation of recurrence equations capturing the evolution of
relevant metrics during execution, and the proof of bounds on the solutions to
these recurrences, often by induction.
Moreover, such analysis must be repeated if the established bound is
 unsatisfactory, after updating the set of metrics or invariants and/or the
algorithm itself. At scale, this workflow can become tedious and time-consuming.

\begin{example*}
  The history of quantifier-elimination procedures for Presburger arithmetic
  (PrA) offers an example of the above picture. Following Oppen's first
  triply-exponential-time upper bound~\cite{Oppen78}, the complexity analysis was
  revisited and refined in a series of papers~\cite{Reddy78,Weispfenning90,Weispfenning97,Haase14}, and
  (prompted by further algorithmic improvements) once more in recent
  work~\cite{ChistikovMS24,MansuttiS25}. All these works follow the blueprint described above, starting
  from the definition of a suitable set of metrics (e.g., the largest bit size
  of an integer occurring in the formula).
\end{example*}

The substantial existing bodies of work in automated cost analysis and symbolic
computation, as well as recent advances, suggest that large parts of the above
workflow are amenable to automation.
We are currently developing a tool with precisely this goal.

In this extended abstract, we sketch the envisioned pipeline, and introduce some
of the theory underlying the tool---including new techniques in recurrence-based static
cost analysis.

\subparagraph*{Challenges.}

The algorithms we consider present interesting difficulties for automated analyses.
Indeed, they require reasoning about highly non-linear, non-polynomial invariants,
out of scope of classical numerical analyses, motivating recurrence-based approaches.
They also involve broad classes of metrics and relations between them,
while making it important to keep extracted recurrences as precise as possible;
both touch on open problems in recurrence-based cost analysis.
Finally, in their case, full automation is unrealistic, which encourages the development of
interactive workflows between theorem provers and automated analysers.

\subparagraph*{Tool architecture.}
The pipeline of our tool, described in the abstract of page~1, is illustrated in Fig.~\ref{fig:pipeline}.
The following sections describe interesting features of each of its main components.

\begin{figure}[H]
  {\scalebox{.85}{
    \centering
    \begin{tikzpicture}[node distance=0.4cm]
        \node[toolstep={1cm}{2.1cm}] (itp)
            {\scalebox{0.8}{\begin{minipage}{3.7cm}
                \centering
                \textbf{\large ITP}\\[-3pt]
                ~\rule{\linewidth}{0.4pt}\\[4pt]
                \setlist[itemize]{leftmargin=*, itemsep=1pt, topsep=0pt}%
                \begin{itemize}
                    \renewcommand{\labelitemi}{\scriptsize$\blacksquare$}
                    \small
                    \item Algorithms
                    \item Metrics $m \colon \Data_T \to \zz$
                    \item Lemmas and proofs
                \end{itemize}
            \end{minipage}}
            };

        \node[toolstep={1cm}{2.1cm}] (ir) [right = 1cm of itp]
          {\scalebox{0.8}{\begin{minipage}{3.5cm}
                \centering
                \textbf{\large IR}\\[-8pt]
                ~\rule{\linewidth}{0.4pt}
                Analysable program:
                \setlist[itemize]{leftmargin=*, itemsep=1pt, topsep=0pt}%
                \begin{itemize}
                    \renewcommand{\labelitemi}{\scriptsize$\blacksquare$}
                    \small
                    \item Algebraic data types
                    \item Catamorphic metrics
                    \item Assumed lemmas
                \end{itemize}
            \end{minipage}}
            };

        \draw[->] (itp) edge node [midway, above] {} (ir);

        \node[toolstep={1cm}{2.1cm}] (receq) [right = 2.1cm of ir]
        {\hspace{3pt}{\scalebox{0.8}{\begin{minipage}{2.1cm}
              \centering
              \textbf{\large Generalised \\ Recurrence \\ Equations }
          \end{minipage}
        }}};

        \draw[->] (ir) edge node [midway, above] {\parbox{2cm}{\centering\footnotesize\textbf{Abstract\\compilation}\\Sec.~\ref{sec:abstract-compilation}}} (receq);

        \node[toolstep={1cm}{2.1cm}] (sols) [right = 2.1cm of receq]
        {\hspace{3pt}{\scalebox{0.8}{\begin{minipage}{2.1cm}
              \centering
              \textbf{\large Parametric Bounds}
          \end{minipage}
        }}};

        \draw[->] (receq) edge node [midway, above] {\parbox{2.25cm}{\centering\footnotesize\textbf{Recurrence\\solving}\\Sec.~\ref{sec:recsolv}}} (sols);

    \end{tikzpicture}
    }}
    \caption{The pipeline of \tool.}\label{fig:pipeline}
\end{figure}

\subparagraph*{Input programs.}

The first step of the tool consists in translating the algorithm to be analysed
from an Interactive Theorem Prover (ITP) into an Intermediate Representation
(IR) that is designed for analysis convenience. Besides the algorithm, the user
provides \emph{metrics}, and optionally \emph{lemmas} directly in the ITP environment.
The auxiliary lemmas are used to improve the analysis; they are \emph{assumed true}
(and their proof erased) in the IR.

At the time of writing this abstract, we are implementing this translation
starting from \texttt{LiquidHaskell}~\cite{Vazou:icfp2014:LiqHaskell-short-nourl} as a frontend. The IR is an imperative language
featuring (first-order, monomorphic) recursive functions, ADTs, metric definitions, and
non-deterministic primitives (\texttt{top}, \texttt{assume}) that reflect
underspecified objects and user lemmas.

\newpage

\section{(Non-)catamorphic metrics and lemmas}
\label{sec:entry-and-metrics}

\subparagraph*{Metrics.}
The central object of the first step of the pipeline is the notion of
\emph{(size) metric}.
\begin{definition}[Size metric] \label{def:metric}
  Given a datatype $T$, a \emph{(size) metric} on $T$ is any
  integer-valued function $m:T\to \zz$.
  For $t\in T$, we say that $m(t)$ is the \emph{size of $t$ (under the metric $m$)}.
\end{definition}
The \emph{recurrences} built in recurrence-based cost analysis relate sizes of program variables at
different program points---e.g., for functions, relations between input and output sizes.
Bounds on the \emph{cost} of a function are then expressed as functions of the
\emph{sizes} of its inputs.\footnote{We only discuss
equations on \emph{sizes}
here; \emph{costs} are similar
after introducing \emph{cost models} (e.g.~\cite{resource-iclp07-shorter-with-doi}).}

Familiar examples include list length, integer value, or tree height: these arise both in elementary
(manual) complexity analysis, and well-known automated complexity analyses. %
While Definition~\ref{def:metric} is extremely general, precise analyses may require more
sophisticated metrics than the examples above.
For example, \emph{potentials} in amortised complexity analysis~\cite{tarjan85} may be seen as instances of size
metrics.
One may also need to consider exotic metrics, such as the total number of elements in a data
structure satisfying some logical property, the \emph{cardinality} of elements of a particular
shape, or number-theoretical properties of integers.

Compared to previous work, our contribution supports fully-automated recurrence extraction for a
broader class of metrics, which we call \emph{catamorphic} (collections of) \emph{metrics}.

\begin{definition}[Catamorphic collection of metrics]

  Consider a collection $\mathcal{T}=(T_1, ..., T_n)$ of algebraic data types (with only base type
  $\zz$), possibly mutually recursive, given categorically as the initial algebra
  $in:F\mathcal{T} \overset{\simeq}{\to}\mathcal{T}$ of some endofunctor $F$ in the $n$-fold
  (cartesian) product $\mathtt{Types}^n$ of the appropriate category of types.
  A collection $\Metrics$ of metrics, composed of $d_i$ metrics of type $T_i\to\zz$ on each $T_i$,
  may be viewed as a morphism $\Metrics:\mathcal{T}\to{\prod_i\zz^{d_i}}$.

  We say that $\Metrics$ is a \emph{catamorphic collection of metrics} whenever there exists a
  morphism $rm: F\big({\prod_i\zz^{d_i}}\big) \to {\prod_i\zz^{d_i}}$ such that $\Metrics \circ
  in = rm \circ F(\Metrics)$.

  Intuitively, $\Metrics$ is catamorphic whenever the size of a term depends only on the sizes of
  the arguments of its constructor.
\end{definition}

\begin{example*}
  Consider the type $\ListInt$ of list of integers, with corresponding functor
  $F:X\to 1 + \zz\times X$, where the two components of the coproduct respectively
  correspond to the $\Nil$ and $\Cons$ constructors.
  The number of distinct integers in a list, $\cardm$, is %
  a \emph{non-}catamorphic metric.
  Indeed, $\cardm(\Cons(\mathtt{hd},\mathtt{tl}))$ does not depend only
  of $\cardm(\mathtt{tl})$ and the value of $\mathtt{hd}$, but also of the
  actual elements of $\mathtt{tl}$.

  In contrast, the \emph{length} %
  of a list is a trivial catamorphic metric,
  where $rm_{\len,\Nil}() = 0$ and $rm_{\len,\Cons}(\_,n) = 1+n$ for $n\in\zz$.
  More subtle catamorphic metrics include the number of elements
  satisfying some arbitrary property or the product of all elements.
\end{example*}

\subparagraph*{Lemmas.}
Our tool also supports user-defined \emph{non-catamorphic metrics}. %
These metrics are automatically translated into \emph{catamorphic overapproximations}, a step that
causes the automated analysis to lose precision.
In this case, we let the user intervene by providing lemmas in the ITP environment,
which are used to recover lost precision.

\begin{example*}
  Consider a program function $\sconcat:\ListInt \to \ListInt$ concatenating a list with itself.
  The tool easily infers that $\len(\sconcat(\ell)) = 2\cdot\len(\ell)$, but, working with catamorphisms
  alone, it can only infer that $\cardm(\sconcat(\ell)) \leq 2\cdot\cardm(\ell)$.

  Were the user to provide a lemma stating $\cardm(\sconcat(\ell))\leq\cardm(\ell)$,
  the tool would exploit this information to sharpen the analysis of all procedures calling $\sconcat$.
\end{example*}

\newpage

\section{Recurrence extraction by abstract compilation}
\label{sec:abstract-compilation}

When extracting recurrences from programs, it is important to ensure that they produce solutions
that are \emph{sound},
and %
it is
desirable that these are \emph{precise}.
Indeed, intuitively, several sets of relations may be inferred between sizes %
at different program points, but different %
sets may provide different information on the concrete behaviour of the program.
Order theory and abstract interpretation~\cite{Cousot77-short-with-doi}
are convenient frameworks to study these notions.

Compared to previous work, we refine the notion of \emph{soundness} of an equation with respect to
the program it abstracts:
our formalisation compares \emph{operators} instead of their fixed points (``solutions'').
This makes equations reflect the \emph{control flow} of the original program.
Moreover, via Galois connections, this provides a satisfying notion of
``\textbf{best} equation that can be extracted from that program'',
under given signatures and sets of metrics,
and a methodology to inductively overapproximate it.

\subsection{Programs and recurrences as operators and their abstractions}

We build upon~\cite{order-recsolv-sas24-and-arxiv-nourl-short}:
(generalised) recurrence equations are represented by operators on the function space of their
unknowns. Similarly, we introduce \emph{operator semantics of programs}, which assigns to a program
$\Prog$ an operator $\semop{\Prog}$, where $\lfp\semop{\Prog}$ is its usual concrete semantics.

\begin{definition}[Operator semantics (sketch)]
  Usual denotational semantics of a program $\Prog$ that recursively defines functions $f\in\Funs$,
  with non-deterministic primitives, can be given as $\lfp\semop{\Prog}$,
  where $\semop{\Prog}\in\End\Big(\prod_{\Funs\ni f:\Pi_i T_i \to T_o}\big({\prod_i T_i\to\pp(T_o)}\big)\Big)$.

  We take the operator $\semop{\Prog}$ as our concrete operator semantics (delaying the fixed point).
\end{definition}
$\semop{\Prog}$ may be viewed as a semantic representation \emph{of the program itself},
control-flow and recursive structure included, with only syntactic features hidden away.
Because of this, Galois connections in operator space %
may be viewed as \emph{abstractions of program themselves}.

\begin{proposition}[\cite{absfun-sttt26-medium-nourl}]
  Let $\Metrics_{T_i}$ be (any) collection of metrics on each $T_i$.
  There is a ``size abstraction'' Galois connection from concrete programs
  (viewed as operators $\semop{\Prog}$ on symbolic functions)
  to (non-deterministic) purely numerical programs, represented as
  $\asemop{\Prog}\in\End\Big(\prod_{f\in\Funs}\big({\prod_i \zz^{|\Metrics_{T_i}|}\to\pp(\zz^{|\Metrics_{T_o}|})}\big)\Big)$.
  These can be further abstracted, e.g.\ to recurrence equations on interval-valued functions (where $\ii$
  replaces $\pp$ above).
  By including $\ii(\zz)$ in $\zz^2$, we recover recurrence equations on usual integer-valued functions.
\end{proposition}

We call \emph{abstract compilation} such transformations (adding back syntax), that turn a concrete
program (e.g.\ symbolic \IR programs) into a simplified program on a new signature (e.g.\ numerical
logic programs, recurrence equations, etc.).
Our contribution solves, for catamorphic metrics, one of the challenges raised
by~\cite{absfun-sttt26-medium-nourl}: the \emph{computation} of such $\asemop{\Prog}$.

\subsection{An abstract compiler for \IR}

We have implemented these ideas in an abstract interpreter for \IR. It contains higher-order
abstract domains, as well as a generic ``size lifter'' (a domain functor) that applies numerical
abstractions to symbolic programs through user-provided catamorphic metrics.

This allows fully automated abstract compilation of \IR programs to precise recurrence equations,
reflecting control flow and piecewise behaviour.

We also support compilation to first-order logic programs over extended integer arithmetic,
which is a more precise representation than interval recurrence equations.
This enables, e.g., reuse of abstract interpreters for logic programs, as well as SMT-based invariant
verification.

\section{Recurrence solving: search and verification of (inductive) bounds}
\label{sec:recsolv}

In favourable cases, the obtained equations may be directly solvable by a
mainstream computer algebra system (CAS) %
or fall within the scope of classical methods such as the master theorem. %
However, this need not be the case in general, and the precise equations
extracted by abstract compilation may not admit any closed-form solution:
indeed, they closely mirror program themselves, which can present complex features.

In such cases, we %
adopt the order-theoretical approach of~\cite{order-recsolv-sas24-and-arxiv-nourl-short}, and search instead for
pre/postfixpoints of the operator $\Phi$, i.e. \emph{inductive} bounds of the solution.

\begin{proposition}[\cite{order-recsolv-sas24-and-arxiv-nourl-short}, Knaster-Tarski corollary]
  Let $\Phi\in\End(\mathfrak{F})$ be a \emph{monotone} equation,
  i.e. an order-preserving operator on the underlying function space
  $(\mathfrak{F},\pleq)$, assumed to be a complete lattice.
  Then, $\lfp\Phi$ exists, and is what we call the \emph{solution} $\fsol$ to the equation.

  For any $\fcand$, $\Phi\fcand \pleq \fcand \implies \fsol \pleq \fcand$,
  i.e. all postfixpoints are upper bounds.

  Moreover, under termination assumptions, we show that $\lfp\Phi = \gfp\Phi$,
  so that we also have $\fcand \pleq \Phi\fcand \implies \fcand \pleq \fsol$.
\end{proposition}

As observed by several recent works~\cite{order-recsolv-sas24-and-arxiv-nourl-short,Goharshady-ESOP2025,absfun-sttt26-medium-nourl,kura2026supermartingalesuniquefixedpoints-nourl},
various pre/postfixpoint search techniques can be applied to discover
inductive bounds. %
These include abstract interpretation,
constrained optimisation, and templates mixed with quantifier elimination, among others.
\subsection{The function comparison problem and transcendental SMT}

Some of these techniques produce \emph{candidate} %
bounds $\fcand$,
and require %
additional work
to \emph{verify} the inductivity condition
$\forall \vec{n}\,\Phi(\fcand)(\vec{n}) \leq \fcand(\vec{n})$.
This is a non-trivial problem,
even in the single-quantifier case %
(\emph{inference} with templates, instead, has one quantifier alternation).

Previous work has mainly focused on \emph{incomplete} techniques based on CAS and rewriting.
However, when $\fcand$ and $\Phi(\fcand)$ are both (piecewise) polynomials, the comparison can be
decided, at least under real relaxation, thanks to the decidability of real polynomial arithmetic.
Moreover, the algorithmic theory of transcendental real and integer arithmetic, e.g.\ extended with
exponentials and logarithms, has seen
significant progress in recent
years~\cite{IncLinNTA-TOCL18,PA-complexity-ICALP23-short-nourl,ChistikovMS24,SWINE-IJCAR24-short,GallegoLM26}.
This motivates us to explore the use of
a recent SMT tool,
\texttt{Yices-TRA}\footnote{\url{https://github.com/arith-lab/yices-tra}},
an extension of the \texttt{Yices} solver supporting nonlinear and
transcendental arithmetic, which we are currently integrating as a
backend to our pipeline.

\subsection{Outlook: integration of termination-analysis techniques}

These approaches suggest several research directions at the interface with termination analysis.

For example, as mentioned above and explored in subtler cases by~\cite{kura2026supermartingalesuniquefixedpoints-nourl}, \emph{termination
hypotheses} on equations themselves are key to obtaining lower bounds on least fixed points.

Termination of equations is also relevant to their \emph{execution}, which is a building block in
some optimisation-based techniques~\cite{mlrec-tplp2024-nourl-short,order-recsolv-sas24-and-arxiv-nourl-short}.
However, even for terminating programs, artificial non-termination may be introduced by
non-determism of abstractions. %
Examples indicate possible
ways to handle these situations, but no general theory is available yet.

Finally, \emph{ranking functions} are central to several complexity
analysers~\cite{Lommen-ESOP26-short}, and support, e.g., the common
counter-based approach to equation setup (see \cite{regular-path-clauses-hcvs21-short-no-url} and related
work).
In a more mysterious sense, they also seem central to the \emph{geometry of sets of postfixpoints}
(see \cite{order-recsolv-sas24-and-arxiv-nourl-short}, Theorem~3), and suggest repair/refine approaches to
function space exploration.

\newpage
\enlargethispage{3\baselineskip}
\bibliography{extended-abstract-bib}

\end{document}